\documentclass[12pt]{article}

\usepackage{textcomp}
\usepackage{chetnew}
\usepackage{mathtools}
\usepackage{amssymb,amsmath,amsfonts,mathrsfs,braket}
\usepackage[utf8]{inputenc}
\usepackage{tikz,tikz-cd,url}
\usepackage{xcolor}
\usetikzlibrary{arrows,decorations.markings,shapes.geometric,decorations.pathmorphing}
\tikzset{snake it/.style={decorate, decoration=snake}}
\usepackage{enumitem}
\usepackage{bm}

\def\one{{\,\hbox{1\kern-.8mm l}}}

\allowdisplaybreaks

\def\makeatletter{\catcode`\@=11}
\makeatletter
\def\mathbox#1{\hbox{$\m@th#1$}}%
\def\math@ccstyles#1#2#3#4#5#6#7{{\leavevmode
      \setbox0\mathbox{#6#7}%
      \setbox2\mathbox{#4#5}%
      \dimen@ #3%
      \baselineskip\z@\lineskiplimit#1\lineskip\z@
      \vbox{\ialign{##\crcr
             \hfil \kern #2\box2 \hfil\crcr
             \noalign{\kern\dimen@}%
             \hfil\box0\hfil\crcr}}}}
\def\mathaccstyles{\math@ccstyles\maxdimen}
\def\maththroughstyles{\math@ccstyles{-\maxdimen}}
\def\unity%
 {\maththroughstyles{.45\ht0}\z@\displaystyle {\mathchar"006C}\displaystyle 1}

\def\FF{{\cal F}}

\def\LL{{\cal L}}

\def\beq{\begin{equation}}
\def\eeq{\end{equation}}
\newcommand{\bea}{\begin{eqnarray}}
\newcommand{\eea}{\end{eqnarray}}
\def\bal{\begin{align}}
\def\eal{\end{align}}

\preprint{CCTP-2024-4 \\ ITCP-IPP-2024/4 \\ QMUL-PH-24-08}

\title{Learning S-Matrix Phases with Neural Operators} 

\author{
Vasilis~Niarchos\;$^{a,\nabla}$ and Constantinos~Papageorgakis\;$^{b,\Box}$
}

\affiliation{
$^a$ ITCP \& CCTP, Department of Physics,\\
University of Crete, 71003 Heraklion, Greece\\
$^b$ Centre for Theoretical Physics, Department of Physics and Astronomy\\ Queen Mary University of London, London E1 4NS, UK \vspace{0.3cm} $ $ \\
\vspace{0.3cm} $ $

\vspace{-1.0cm}
{\tt \small
$^\nabla$niarchos@physics.uoc.gr,  
$^\Box$c.papageorgakis@qmul.ac.uk}}

\abstract{We use Fourier Neural Operators (FNOs) to study the relation between the modulus and phase of amplitudes in $2\to 2$ elastic scattering at fixed energies. Unlike previous approaches, we do not employ the integral relation imposed by unitarity, but instead train FNOs to discover it from many samples of amplitudes with finite partial wave expansions. When trained only on true samples, the FNO correctly predicts (unique or ambiguous) phases of amplitudes with infinite partial wave expansions. When also trained on false samples, it can rate the quality of its prediction by producing a true/false classifying index. We observe that the value of this index is strongly correlated with the violation of the unitarity constraint for the predicted phase, and present examples where it delineates the boundary between allowed and disallowed profiles of the modulus. Our application of FNOs is unconventional: it involves a simultaneous regression-classification task and emphasizes the role of statistics in ensembles of NOs. We comment on the merits and limitations of the approach and its potential as a new methodology in Theoretical Physics.}

\date{\today}

\begin{document}

\maketitle

\hypersetup{pageanchor=true}
\setcounter{tocdepth}{2}

\toc


\section{Introduction}
\label{intro}

The vast majority of problems in Physics and Mathematics involve the study of different types of functional relations. On general terms, these relations can be viewed as maps between infinite-dimensional spaces of functions. Sometimes, the origin of these maps is well understood. For example, a function may be obtained as the solution to an integral or differential equation that involves other input functions (e.g.\ functions that specify the form of the equation, boundary conditions, etc.). Analytic solutions are usually tractable only in special cases, while generic situations are computationally hard and require approximate schemes and numerical methods. 

There are also many contexts where the rules dictating the map of interest are either poorly understood or beyond the reach of the existing framework. This is common for interacting, non-perturbative Quantum Field Theories (QFTs). For example, in QFTs with a standard Lagrangian formulation, one would like to understand the map between spacetime-dependent deformations of the action by arbitrary operators, expressed by source functions in spacetime, and the partition function of the theory (or its functional derivative with respect to the sources). The partition function contains all the necessary information about the local correlation functions of the QFT, which are some of the main objects of interest in quantum Physics. The traditional computation of the partition function goes through a path integral, which is typically difficult to evaluate, and in many cases also difficult to properly define.\footnote{Note that in the (super)gravity limit of the AdS/CFT correspondence \cite{Maldacena:1997re} the map between sources and functional derivatives of the partition function reduces to the solution of Partial Differential Equations in classical gravity with suitable boundary conditions. This translates the QFT problem back to the study of functional relations in the context of differential equations mentioned above.} 

In recent years, investigations originating from String Theory have also revealed many new examples of QFTs that do not seem to admit a Lagrangian formulation and therefore challenge the traditional Lagrangian and Hamiltonian framework of quantum theories. There is very little we can currently compute in such theories with existing methods. This fact has motivated a flurry of activity in search of new non-perturbative approaches to QFTs. The modern Conformal bootstrap and $S$-matrix bootstrap programs \cite{Poland:2018epd,Rychkov:2023wsd,Kruczenski:2022lot} are prominent examples.

For the above reasons, it is particularly interesting to develop novel methodologies that will allow us to better understand general maps between functions in various contexts. We are especially interested in situations where partial information from explicit solutions in special tractable cases can be used to uncover hidden structures, and achieve generalizations towards computationally hard generic regimes. Can data-driven methods help in this direction? Can they produce reliable results with quantifiable error, and potentially new analytic understanding?

In this paper, we would like to probe these general questions in a very specific problem, that concerns the relation between the modulus and the phase of scattering amplitudes in elastic $2\to 2$ scattering at fixed energies. This relation, which is an important ingredient of $S$-matrix theory, is constrained by unitarity through a non-trivial integral equation (see Eq.\ \eqref{unitarity} below). Instead of solving this equation directly, we will attempt to re-discover it by ``learning'' it from the data of amplitudes with finite partial wave expansions, where both the modulus and phase are straightforward to compute as functions of real phase shifts. 

We will study the relationship between modulus and phase (and the implications of unitarity) using a modern supervised Machine Learning technique: {\it Neural Operators}  (NOs) \cite{kovachki2023neural,li2020fourier}. Unlike standard Neural Networks that are good function approximators, Neural Operators are good approximators of maps between infinite-dimensional function spaces. Since we are seeking to learn the map between the modulus and phase of a scattering amplitude---both functions of the scattering angle---NOs present themselves as an appealing tool. 

Our main goal in this context will be to explore:
\begin{enumerate}[label=({\it \roman*}\,)]
    \item 
    To what extent NOs generalize knowledge from finite to infinite partial wave expansions.

    \item 
    How to quantify the reliability of the result assuming no prior knowledge of the unitarity constraint.
\end{enumerate}

Towards that end, we will run a simultaneous {\it regression-classification task} by training the NOs on both true and false samples. Their output will contain an extra label, which will be called ``fidelity index'', indicating whether the prediction should be kept as a reliable solution or get rejected. We will provide evidence that the fidelity index extracts non-trivial features of the true solutions and that its value correlates with the degree of violation of the unitarity equation. 

Typically, NOs supplement other direct methods in the solution of complicated equations, commonly Partial Differential Equations (PDEs). The above implementation of NOs in a simultaneous regression-classification task is unconventional; to the best of our knowledge similar applications have been thus far limited  
(for some recent studies of NOs in image classification see \cite{9820128,rs14122931,kabri2023resolution,kashefi2024novel}). 

The performance of a NO---and how it learns---for fixed hyperparameters and training datasets depends on various stochastic factors that play a role during the training process and are hard to quantify. We will therefore also propose that it is useful to study the {\it collective} behavior of NOs. In particular, we will present specific data exhibiting the improved properties of the {\it mean fidelity index}. We will argue that quantities like the mean fidelity index can be useful, and could play a role similar to the Martin parameter $\sin\mu$ (see Eq.\ \eqref{martin}), which provides a partial characterization of the scattering amplitudes. 
    
Setting NOs aside for a moment, another popular Machine Learning method that appears in the context of PDEs is Physics Informed Neural Networks (PINNs)\cite{raissi}. In that case, Neural Networks are used to directly model the unknown function: the equation to be solved goes into the definition of the loss that the network tries to minimize during training. Recently, PINNs were used to directly solve for the unitarity equation, obtaining notable results \cite{Dersy:2023job}. We emphasize that our approach should be viewed as complementary with an orthogonal scope, because we are attempting to reconstruct the unitarity equation and its implications without using it directly.   

The rest of this paper is organized as follows. We begin in Section~\ref{background} with an introduction of the Physics problem and a summary of the key formulas and definitions used in the main text. In Section~\ref{NNs} we present the salient features of PINNs and NOs, along with useful references for the non-expert reader. The main results of the paper appear in Section~\ref{single}, that focuses on amplitudes with unique phases, and Section \ref{double}, that discusses the subtle case of amplitudes with phase ambiguities. In both cases, we see that NOs can generalize non-trivially beyond their training set, learning important properties about the structure of the system. We elaborate on the efficiency, advantages, disadvantages and difficulties of the approach. We conclude in Section \ref{conclusions} with a brief summary of our main results and a discussion of interesting future prospects.

\section{Background: Modulus and Phase in Elastic Scattering}\label{background}

The following discussion is restricted to elastic $2\to 2$ scattering. In quantum scattering processes, we measure the differential cross-section $\frac{d\sigma}{d\Omega}$, which is equal to the square of the modulus of the scattering amplitude $f(\theta)$:
\beq
\label{backaa}
\frac{d\sigma}{d\Omega} = |f(\theta)|^2
~.
\eeq
The scattering amplitude, which is part of the asymptotic form of the wavefunction in non-relativistic Quantum Mechanics, is a complex number
\beq
\label{backab}
f(z) = b(z)\, e^{i\phi(z)}
~,
\eeq
with modulus $b(z)$ and phase $\phi(z)$. We used $z:=\cos\theta$ to express the dependence on the scattering angle $\theta$. From the differential scattering cross-section one reads off $b(z)$, but it is in principle difficult to extract the corresponding phase $\phi(z)$. 

Mathematically, this task is easy when the scattering amplitude admits a {\it finite} partial wave expansion
\beq
\label{backac}
f(z) = \frac{1}{k} \sum_{\ell=0}^L (2\ell+1) \sin\delta_\ell e^{i\delta_\ell} P_\ell(z)
\eeq
in terms of $L+1$ phase-shifts $\delta_\ell$. Both $b(z)$ and $\phi(z)$ are expressed in terms of the phase-shifts. In this form, unitarity plays a simple role; it dictates that the phase-shifts are real. In \eqref{backac}, $k$ is the wavenumber of a non-relativistic particle scattered by some potential in Quantum Mechanics and the $P_\ell(z)$ are Legendre polynomials. 

A generic amplitude, however, admits an ${\it infinite}$ partial wave expansion. At fixed energy (equivalently, fixed $k$) the rescaled amplitude $F(z)=k f(z)$ is an infinite superposition of partial waves
\beq
\label{backad}
F(z)=B(z) e^{i \phi(z)}=\sum_{\ell=0}^{\infty}(2 \ell+1) \sin{\delta_\ell}e^{i \delta_\ell} P_{\ell}(z)
~.
\eeq
In that case, finding the phase $\phi(z)$ for a given $B(z)$ is more complicated and the partial wave expansion is less useful.

Nevertheless, when formulated more generally, unitarity is a strong condition that non-trivially relates the modulus and the phase of a scattering amplitude. A standard argument\footnote{See \cite{landau1981quantum, RevModPhys.46.369} for a review of the argument in non-relativistic Quantum Mechanics and \cite{Martin:1969ina,Correia:2020xtr} for a discussion in relativistic QFT. A related discussion also appears in \cite{Newton:1968zs, Atkinson:1970zza}.} shows that unitarity imposes the integral constraint
\begin{align}\label{unitarity}
\sin \phi(z)=\int_{-1}^1 d z_1 \int_0^{2 \pi} d \phi_1 \frac{B\left(z_1\right) B\left(z_2\right)}{4 \pi B(z)} \cos \left[\phi\left(z_1\right)-\phi\left(z_2\right)\right]\;,
\end{align}
where
\begin{align}
\label{z2}
    z_2\left(z, z_1, \phi_1\right) \equiv z z_1+\sqrt{1-z^2} \sqrt{1-z_1^2} \cos \phi_1\;.
\end{align}
For a given $B(z)$, one would like to solve this equation to determine the corresponding phase $\phi(z)$. 

In the existing literature, a significant amount of effort has been put into determining for which $B(z)$ there exist solutions for $\phi(z)$, either unique or multiple, and several associated bounds have been established. The so-called ``dual bound'' is derived by setting $z=1$ in \eqref{unitarity}. This special case provides a necessary condition for $B(z)$ to be the valid modulus of a scattering amplitude:
\begin{align}\label{dualbound}
\int_{-1}^1 d z_1 \frac{B\left(z_1\right)^2}{2 B(1)} \leq 1\;.
\end{align}
Additional bounds on existence and uniqueness can be obtained by defining the function
\begin{align}
    K(z) := \int_{-1}^1 d z_1 \int_0^{2 \pi} d \phi_1 \frac{B\left(z_1\right) B\left(z_2\right)}{4 \pi B(z)}
\end{align}
and the ``Martin parameter''
\begin{align}
\label{martin}
    \sin\mu:= \max_{-1\le z\le 1} K(z)\;.
\end{align}
For example, one can trivially show using \eqref{unitarity} that
\begin{align}
|\sin \phi(z)| \leq \sin \mu\;.
\end{align}
Moreover it can be proven that, given a modulus $B(z)$, solutions for phases always exist when $\sin\mu\le 1$ \cite{Martin:1969xs} but known arguments do not preclude the existence of solutions also for $\sin\mu>1$. Polynomial (finite partial wave) amplitudes are unique if $\sin\mu\le 1$ \cite{Martin:1969xs}. For amplitudes with an infinite number of partial waves the best bound on uniqueness is currently $\sin\mu <0.86 $ \cite{Gangal:1983uy}, but it is believed that phases should be unique up to  $\sin\mu<1$, \cite{Martin:1969xs,Atkinson:1970zza}.

There can also be multiple (ambiguous) phases corresponding to the same modulus, which do not include the trivial ambiguity where all the $\delta_\ell\to -\delta_\ell$ (and, therefore, $F(z)\to -F(z)^*$ via \eqref{backad}). For elastic scattering this degeneracy is two-fold \cite{Itzykson:1973sz,Martin:2020jlu} and has been completely classified for finite partial waves with $L=2,3$ \cite{Crichton:66,Atkinson:1973je,Berends:1973mg}. Phase ambiguities in $L=4$ amplitudes have been discussed in \cite{Cornille:1974zp}. Two-fold ambiguous solutions can also be constructed for amplitudes with infinite partial wave expansions. It is interesting to ask what is the lowest possible value of $\sin\mu$ for the ambiguous solutions. For example, for $L=2$ the lowest value of $\sin\mu$ is 2.6. An amplitude with the lowest known value of $\sin\mu \simeq 1.67$ was constructed recently using Machine Learning methods in \cite{Dersy:2023job}.

\section{PINNs, Neural Operators and PINOs}\label{NNs}

We next summarize some of the high-level features of Physics Informed Neural Networks (PINNs) and Neural Operators for 
the non-expert reader, and highlight their main differences. 

Let us assume that we want to solve a system of equations for a set of unknown functions. In many applications, this is a system of Partial Differential Equations (PDEs), or a system of integro-differential equations, or a set of algebraic equations. A natural Machine Learning approach is to use Neural Networks (NNs) as universal function approximators \cite{Cybenko1989ApproximationBS} to model the unknown functions and set up a training process where the parameters of the NNs are optimized to satisfy the prescribed system of equations with the least possible error.\footnote{This process involves the solution of a typically very high-dimensional non-linear, non-convex optimization problem with thousands, millions, or more, parameters. Stochastic Gradient Descent methods have proved very efficient in this context and algorithms like the ADAptive Momentum estimation (ADAM) \cite{Kingma2015AdamAM} are popular choices.} The domain of the functions is discretized on a collocation grid, and the corresponding error in the equations is evaluated and quantified in a scalar semipositive quantity, typically the Mean Squared Error on the grid. This idea forms the basis behind PINNs \cite{raissi} (related ideas go back to several papers from the 1990s, e.g.\ \cite{chenchen1,Lagaris:1997at}) and constitutes an ``unsupervised'' approach: the algorithm generates its own data and tries to solve a problem associated with the specific system of equations. When the form of the equations changes (e.g.\ the source function in a PDE, or the functions that describe the boundary/initial conditions), the PINN needs to be optimized from scratch.

Neural Operators are another data-driven approach that employs NNs. In this case, the idea is to approximate the ``solution operator'' that maps the input functions (e.g.\ source functions, boundary/initial conditions) to the output functions solving the system of equations. To achieve this goal a NN with a more complicated architecture is employed. The latter is not merely the composition of linear operations and point-wise non-linear actions of activation functions, but also convolutions that act non-diagonally on the domain of the input functions. Early discussions of Neural Operators (and related universal approximation theorems) also go back to the 1990s, e.g.\ \cite{Chen1995UniversalAT}. In the present work, we will be employing a modern incarnation of the Neural Operator concept, the so-called Fourier Neural Operators (FNOs), which are constructed using convolution kernels defined in Fourier space \cite{kovachki2023neural,li2020fourier}. Another approach that shares some common features with Neural Operators are the Deep Operator Networks (DeepONets), \cite{Lu2019LearningNO}. We will not consider DeepONets in this paper.

The NO is a ``supervised'' Machine Learning method. The training is based on a dataset of ground-truth input-output pairs, 
that teach the algorithm to map between the input and output function spaces. In typical applications, this dataset is generated by solving the system of equations of interest through some other method. It is worth noting, that although functions are defined on a grid during this process, NOs are discretization invariant and exhibit advanced performance in zero-shot super-resolution---namely, they can be trained at low-resolution samples and compute at never-before-seen high-resolutions \cite{kovachki2023neural,li2020fourier}. Another obvious characteristic advantage of NOs is that, once trained, they can quickly find the solution for new inputs without further re-training, in contradistinction with PINNs. This is convenient if one scans over a landscape of input functions (as we will be doing later in this paper). 

There is a plethora of applications of NOs to PDEs in the literature. A recent application of NOs to the time-dependent Schr\"odinger equation and scattering in non-relativistic Quantum Mechanics appeared in \cite{Mizera:2023bsw}.

Recently, the authors of Ref.\ \cite{Dersy:2023job} employed the PINN approach to study the relation between the modulus and phase of the scattering amplitude in elastic $2\to 2$ scattering, solving the unitarity equation \eqref{unitarity}. They produced remarkable results, including a new solution with ambiguous phases that has the lowest known Martin parameter $\sin\mu\simeq 1.67$. This result improved the relevant bound for the first time in 50 years.

In this paper we do not want to simply repeat the analysis of \cite{Dersy:2023job} using NOs as an alternative Machine Learning method. For the reasons outlined in the introduction, our main motivation is to explore to what extent we can learn the solutions together with the equation we are trying to solve. In the present work, that means learning the modulus/phase relation in Section \ref{background} {\it without} using the unitarity equation \eqref{unitarity}. In this quest, we will be using the NOs in a rather unorthodox way. The NO will be trained on both true and false samples in a class of input functions where \eqref{unitarity} is trivially satisfied, and will be asked to uncover non-trivial structure underlying \eqref{unitarity} outside this class rating its own performance and the quality of its predictions. We hope that this application will inspire other similar explorations in even more complicated problems, where the underlying equations are missing.

As a final comment, we would like to point out that it is also possible to combine the benefits of PINNs and NOs in a hybrid construction that trains NOs using the loss of the underlying equation like a PINN. This approach is called a Physics Informed Neural Operator (PINO) and has been explored in the context of PDEs in \cite{li2023physicsinformed}. It would be interesting to explore potential improvements of the results in this work and \cite{Dersy:2023job} using PINOs.\footnote{It would also be interesting to explore further related applications in the context of the S-matrix bootstrap, see e.g. \cite{Bhat:2023puy}.}

\section{Unique Phases}
\label{single}

In this section, we train a NO on a set of random finite partial wave expansions with $L=1,2,3$ to learn the mapping between the input modulus $B(z)$ and the output $\sin\phi(z)$, the sine of the corresponding amplitude phase. We assume that the relation is one-to-one and set up the training accordingly. Once trained,  we test how well the NO predicts the phase of unseen amplitudes, e.g.\ amplitudes with an infinite partial wave expansion. We also explore  ways to detect whether or not the prediction is reliable.

\subsection{Neural Operator Setup I: Training on Samples of Valid Solutions}
\label{noset1}

We now present our first attempt at NO training. We begin by listing the hyperparameters used and detail the choice of training and test datasets, before testing for generalizations of the trained model. All the computations in this work were performed on NVIDIA A100 GPUs with 40GB RAM. 

\subsubsection{Hyperparameters and Training} 
\label{hyperparams1}

\paragraph{Hyperparameters.} Using the Fourier Neural Operator implementation of \cite{li2020fourier}, for which a well-explained documentation can be found on \href{https://github.com/neuraloperator/neuraloperator}{GitHub}, we set up a 1D Tensorized Fourier Neural Operator (TFNO) implemented in PyTorch with the following hyperparameters:  
\bea
\label{noset1aa}
{\rm number~of~Fourier~modes}~&:& ~~ {\tt n\_ modes}=50~,
\nonumber\\ 
{\rm number~of~hidden~channels}~&:& ~~ {\tt hidden\_ channels}=64~,
\nonumber\\
{\rm number~of~projection~channels}~&:& ~~ {\tt projection\_ channels}=512~,
\nonumber\\
{\rm number~of~layers}~&:& ~~ {\tt n\_ layers}=4~,
\nonumber\\
{\rm type~of~factorization}~&:& ~~ {\tt factorization}={\tt `tucker'}~,
\nonumber\\
{\rm rank}~&:& ~~ {\tt rank}=0.01
~.\nonumber
\eea
This is a model with $76,849$ parameters that are tuned during the training to produce an optimal NO. The training optimization was performed using {\tt ADAM} \cite{Kingma:2014vow} with learning rate $10^{-3}$, weight decay $10^{-5}$ and batch size 256. Varying the above hyperparameters did not result in significant variations of the results.

\paragraph{Training.}
The training dataset is prepared in the following manner. We generate random samples of amplitudes with finite partial wave expansions
\beq
\label{noset1ab}
F(z)=B(z) e^{i \phi(z)}=\sum_{\ell=0}^{L}(2 \ell+1) \sin{\delta_\ell}e^{i \delta_\ell} P_{\ell}(z)
\eeq
sampling the random phase-shifts $\delta_\ell$ from a uniform distribution. 
100K samples are collected for $L=1,2$ and 3, separately, providing a total of 300K amplitudes. For each of these amplitudes we read off their modulus $B(z)$ and the sine of their phase $\sin\phi(z)$. Afterward, we discretize $z=\cos\theta \in [-1,1]$ on a uniform grid of 100 points\footnote{A remarkable feature of NOs is their capacity to efficiently implement zero-shot super-resolution \cite{shocher2017zeroshot}. In the context of quantum $2\to 2$ scattering, this gives us the ability to train on a grid of, say, 100 points and then easily make accurate predictions at higher resolutions. We did not see the need to go beyond the 100-point resolution in this problem, but it is good to keep in mind that this possibility exists.} to produce 300K 100-dimensional vectors $\vec B$ and 300K 100-dimensional vectors $\overrightarrow {\sin\phi}$. The collection of $\vec B$ vectors is converted to a PyTorch tensor that forms the input of the NO during training. Similarly, the collection of $\overrightarrow {\sin\phi}$ vectors is converted to a PyTorch tensor that forms the ground-truth output of the NO. We train on 98\% of the samples (namely, 294K samples) and test on 2\% (namely 6K samples, evenly distributed across $L=1,2,3$). The results reported below are based on a single training run of 6.5K epochs. 

We emphasize that once the trained NO has been obtained, it can be used to make very quick predictions for any input modulus $B(z)$, in stark contrast with the PINN approach, where retraining from scratch is needed for every new input.

\begin{figure}[t!]
    \begin{center}
        \includegraphics[width=0.8\textwidth]{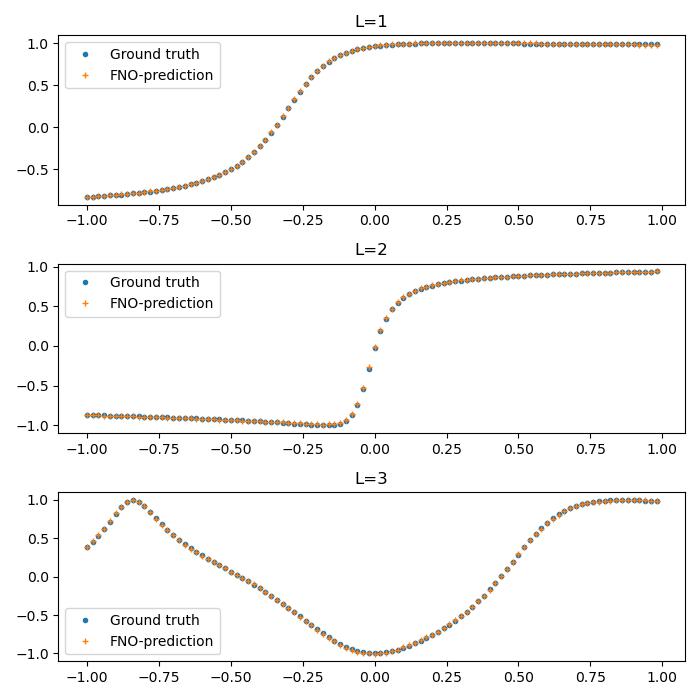}
    \end{center}
    \caption{\small Plots of the ground truth $\sin\phi(z)$ (blue color) and FNO-predicted $\sin\phi(z)$ (orange color) for 3 randomly chosen samples of amplitudes within the 6K test-dataset. From top to bottom we list plots for amplitudes with finite partial wave expansions and $L=1,2,3$ respectively.}
    \label{fig:test_set}
\end{figure}

\subsubsection{Tests against Known Results}
\label{test1}

Once the NO has trained on known samples, we investigate how well it generalizes both on the same class of data (training/test dataset), as well as on different classes of never-before-seen data. Predicting the phases of amplitudes in the latter case would indicate that the NO is able to learn the unitarity relation \eqref{unitarity} and effectively solve it without having direct access to it.

\paragraph{Tests within the training-test dataset.}
For starters, we can ask about the quality of predictions inside the training/test dataset. In Fig.\ \ref{fig:test_set} we plot the ground-truth (blue) and predictions (orange) of the trained NO for $\sin\phi(z)$ of 3 randomly chosen samples from the test dataset with $L=1,2,3$, respectively. The plots of the ground-truth and prediction are visually indistinguishable, indicating that the NO has trained well. To get a sense of the numerical size of the error in the plots of Fig.\ \ref{fig:test_set}, for $L=1$ the average relative error between the ground-truth and prediction across the whole $z$-grid is 0.4\%. For $L=2$ it is 1.1\% and for $L=3$ it is 0.9\%. These numbers are typical in the test dataset. The percentage of samples that exhibit average relative error above 10\% is 5.2\%.  

\begin{figure}[t!]
    \begin{center}
        \includegraphics[width=0.45\textwidth]{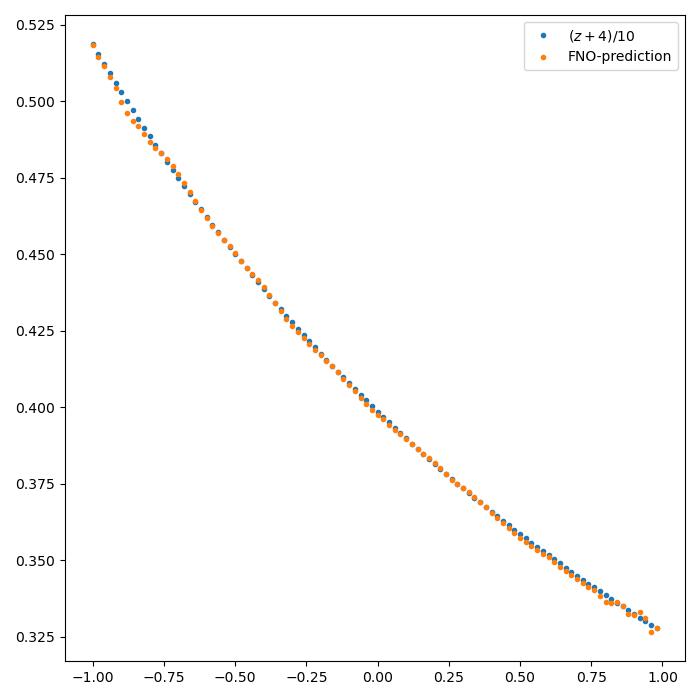}
        \includegraphics[width=0.45\textwidth]{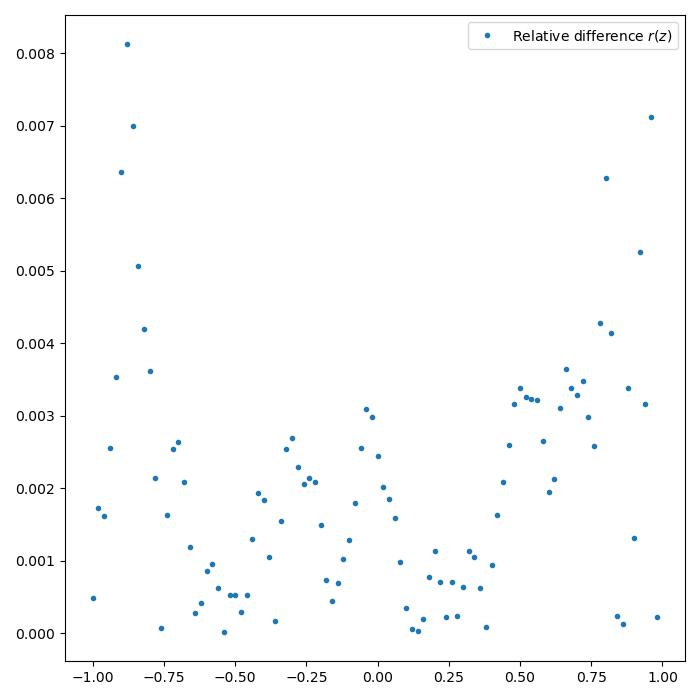}
        \includegraphics[width=0.45\textwidth]{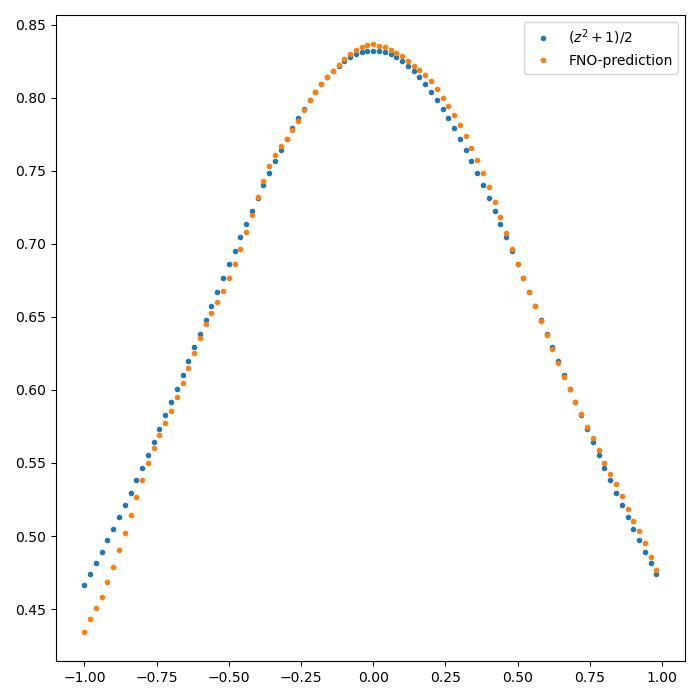}
        \includegraphics[width=0.45\textwidth]{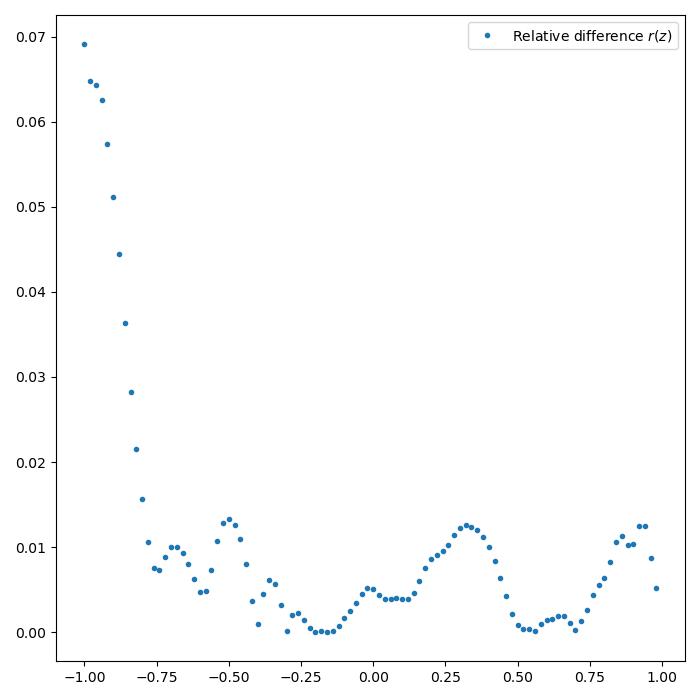}
    \end{center}
    \caption{\small The top two plots display the prediction of the trained NO for $\sin\phi(z)$ against the exact result for input modulus $B(z) = \frac{1}{10}(z+4)$. On the left are the actual functions, while on the right the point-wise relative difference. The bottom two plots display the corresponding data for input modulus $B(z) = \frac{1}{2}(z^2+1)$. Both cases refer to amplitudes with an infinite partial wave expansion.}
    \label{fig:single_predictions}
\end{figure}

\paragraph{A first sample of tests on moduli with infinite partial wave expansions.}
A more interesting question concerns the extent to which the NO can generalize outside the training dataset. The first case we would like to discuss here are amplitudes with an {\it infinite} partial wave expansion. For concreteness, we will consider two examples of linear and quadratic moduli that were analyzed also in Ref.\ \cite{Dersy:2023job}. Later we will scan more extensively over the predictions for amplitudes with linear, quadratic, as well as cubic moduli. 

The first example concerns amplitudes with linear modulus $B(z)=az+b$ (and $b>|a|$ for positivity). It is straightforward to check that these amplitudes do not have a finite partial wave expansion.\footnote{For a detailed discussion see \cite{Dersy:2023job}.} In the top left plot of Fig.\ \ref{fig:single_predictions} we present the prediction of the NO for $\sin\phi(z)$ when $B(z)=\frac{1}{10}(z+4)$ against a numerical solution of the unitarity equation \eqref{unitarity} obtained with the use of an iteration scheme. This particular amplitude has Martin parameter $\sin\mu = 0.522$ and the iteration scheme converges very quickly. The NO prediction is denoted by orange, while the solution of the unitarity equation by blue. The two solutions are visibly close. On the top right plot of Fig.\ \ref{fig:single_predictions} we also present (point-by-point on our $z$-grid) the relative difference $r(z)$ between the NO prediction $\sin\phi_{\rm NO}(z)$ and the solution of the unitarity equation $\sin\phi_{\eqref{unitarity}}(z)$
\beq
r(z) := \left| \frac{\sin\phi_{\eqref{unitarity}}(z) - \sin\phi_{\rm NO}(z)}{\sin\phi_{\eqref{unitarity}}(z)}     
\right|
~.
\eeq
For most points the relative difference is of the order of $O(10^{-3})$.

The second example refers to the quadratic modulus $B(z)=\frac{1}{2}(z^2+1)$, which was also discussed in Ref.\ \cite{Dersy:2023job}. This amplitude has Martin parameter $\sin\mu = 0.867$ and can once again be determined numerically by solving the unitarity equation \eqref{unitarity} with a simple iteration scheme. In the bottom left plot of Fig. \ref{fig:single_predictions} we present in orange and blue, respectively, the $\sin\phi$ for the NO prediction and the solution of the unitarity equation. Once again, the two plots are visibly close. In the bottom right plot of Fig. \ref{fig:single_predictions} we also present the relative difference, which is now of the order of $O(10^{-2})$ for most points. It increases near $z=-1$, where the prediction in the depicted run was less accurate.

\begin{figure}[t!]
    \begin{center}
        \includegraphics[width=0.495\textwidth]{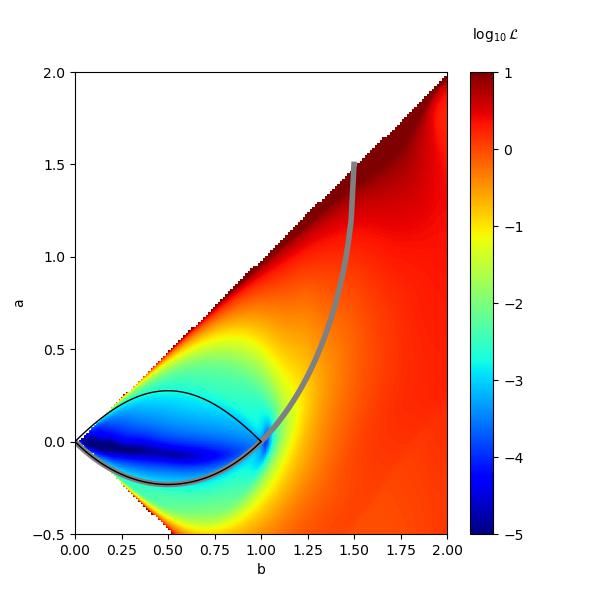}
        \includegraphics[width=0.495\textwidth]{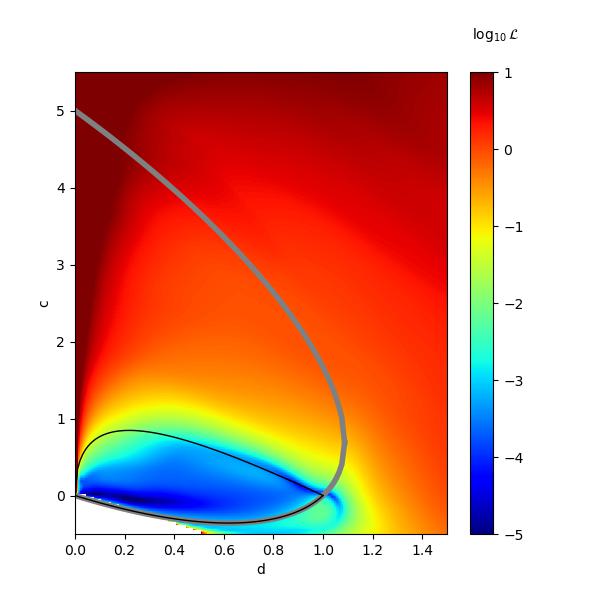}
        \includegraphics[width=0.495\textwidth]{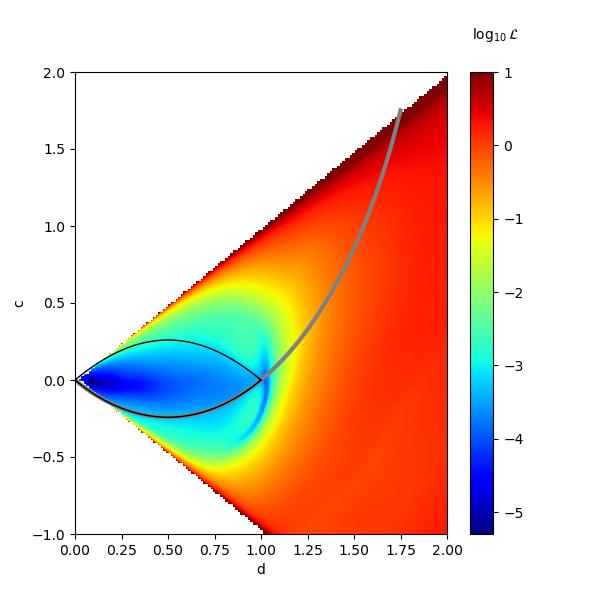}
    \end{center}
    \caption{\small Heatmaps for the log base 10 loss of the NO prediction with respect to the unitarity condition \eqref{unitarity}. The top left plot refers to linear moduli $B(z)=az+b$, the top right plot to quadratic moduli $B(z) = cz^2+d$ and the bottom plot to cubic moduli $B(z)=cz^3+d$. Analogous results for the top two plots were obtained with the use of PINNs in Ref.\ \cite{Dersy:2023job} (see Figs.\ 3 and 5 of that paper). The thin black curves are the $\sin\mu=1$ boundaries while the thick gray curves express the dual bounds.}
    \label{fig:heatmaps}
\end{figure}

\paragraph{Scans on linear, quadratic and cubic moduli.}

We can test the quality of the NO predictions on moduli with infinite partial wave expansions more extensively, by performing a scan over a wide grid of linear, quadratic and cubic moduli. To quantify the quality of the predictions we compute the loss in the unitarity condition \eqref{unitarity}
\beq
\label{test1aa}
\LL := \frac{1}{N_c} \sum_{z_i} \left( 
\sin\phi(z_i) - \frac{1}{4\pi B(z_i)} \int_{-1}^1 dz_1 \int_0^{2\pi} d\phi_1 B(z_1) B(z_{2i}) \cos(\phi(z_1) - \phi(z_{2i})) \right)^2
~,
\eeq
where $z_{2i}$ is computed as in \eqref{z2} with $z=z_i$. The sum is over the points $z_i$ of the collocation $z$-grid and the average is obtained by dividing with the number $N_c$ of collocation points. In our runs $N_c=100$. The integrals in \eqref{test1aa} were computed numerically using the trapezoidal rule.\footnote{The results presented in this section used the fixed grid of 100 collocation points in the NO training in order to apply the trapezoidal rule. It is straightforward to achieve higher numerical accuracy in the numerical computation of the integrals in \eqref{test1aa} using higher resolution grids with NO zero-shot super-resolution.} Fig.\ \ref{fig:heatmaps} displays the heatmaps of the values of $\log_{10}\LL$ for the NO predictions on a grid of linear, quadratic and cubic moduli. Let us comment on each of these plots separately.

For the linear moduli $B(z)=az+b$ we considered a grid of $180 \times 150$ points on the $(a,b)$ plane for $a\in [-0.5,2]$, $b\in [0,2]$ and $b>|a|$. The heatmap of the $\log_{10}\LL$ values on this grid appears on the top left plot of Fig.\ \ref{fig:heatmaps}. The corresponding heatmap in Ref.\ \cite{Dersy:2023job} appears in Fig.\ 3 of that paper. Ref.\ \cite{Dersy:2023job} computed on a grid of $75 \times 60$ points in the $(a,b)$ plane, {\it retraining} the Neural Networks for 2K epochs to obtain each point. Instead, we are {\it evaluating} the already trained NO at each point producing a heatmap on a finer grid within approximately 20 seconds. 

In the top left plot of Fig.\ \ref{fig:heatmaps} we observe a distinct, blue-coloured, low-loss region inside the $\sin\mu=1$ contour, precisely like the one detected by PINNs for linear moduli in \cite{Dersy:2023job}. The main difference with the PINN result is that its lowest $\log_{10}$ losses are in the vicinity of $-8$, whereas our corresponding values are in the vicinity of $-5$. That amounts to a difference in loss between the two methods at the level of 3 orders of magnitude. This is expected, since the PINN performs a dedicated optimization search for each input modulus $B(z)$ explicitly using the unitarity condition \eqref{unitarity}, whereas the NO trains on a completely different class of inputs to produce a prediction outside its training dataset without using the unitarity condition. In that sense, the NO results in Fig.\ \ref{fig:heatmaps} are impressive and provide a distinct indication that the NO has been able to generalize well within the infinite partial wave amplitudes with linear moduli. Of course, by simply looking at the heatmap of Fig.\ \ref{fig:heatmaps} one cannot really deduce where one should put the cutoff that separates the predictions that are consistent with unitarity from the ones that are inconsistent with unitarity. The same issue also exists within the PINN approach, but there it is slightly mitigated by the lower losses of the corresponding results. We will have to say more about how to address this difficulty in the next subsection.   

For the quadratic moduli $B(z)=cz^2+d$ we also considered a grid of $180 \times 150$ points on the $(c,d)$ plane for $c\in[-0.5,,5.5]$, $d\in [0,1.5]$ and $c>|d|$. The $\log_{10}\LL$ heatmap on this grid is depicted on the top right plot of Fig.\ \ref{fig:heatmaps}. The corresponding heatmap from Ref.\ \cite{Dersy:2023job} appears in Fig.\ 5 of that paper. Once again, we observe the formation of a low-loss region inside the $\sin\mu=1$ contour, which is comparable with the result in Fig.\ 5 of Ref.\ \cite{Dersy:2023job}, suggesting that the NO has been able to generalize to this class of amplitudes as well. Similar to the linear case, the NO losses are higher by roughly 3 orders of magnitude compared to the PINN losses of \cite{Dersy:2023job}. 

Finally, in the bottom plot of Fig.\ \ref{fig:heatmaps} we present the $\log_{10}$ unitarity loss for cubic moduli of the form $B(z)=cz^3+d$. Such amplitudes were not discussed in Ref.\ \cite{Dersy:2023job}. The resulting heatmap is comparable to the linear-moduli heatmap, exhibiting a distinct low-loss region inside the $\sin\mu=1$ contour, as expected.

To summarize, in all three cases of infinite partial wave amplitudes analysed in this subsection, the picture that emerges is impressively consistent with expectations from the analysis of the unitarity equation \eqref{unitarity}, suggesting that the NO has learned non-trivial features of that equation without having access to it. The results also exhibit some of the weaknesses of the approach: 
\begin{itemize}
    \item[$(a)$] The lowest losses are a few orders of magnitude higher than those produced by PINNs. That makes it harder to detect, without prior knowledge, the boundary between valid predictions consistent with unitarity and invalid predictions inconsistent with unitarity.  
    \item[$(b)$] The low-loss regions (in blue) are not perfectly aligned with the $\sin\mu=1$ and dual bounds. For example, there are small blue regions violating the dual bounds. 
    \item[$(c)$] The quadratic-modulus NO heatmap in Fig.\ \ref{fig:heatmaps} does not appear to detect the additional solutions appearing in Fig.\ 5 of \cite{Dersy:2023job} (e.g.\ the two small islands of $L=2$ finite-partial wave solutions). 
    \item[$(d)$] To demonstrate non-trivial learning in the above cases, we had to use the unitarity relation \eqref{unitarity}. Without explicit knowledge of that equation the heatmaps in Fig.\ \ref{fig:heatmaps} would not have been possible. In addition, it is unclear to what degree of generality the NO has been able to learn the unitarity equation, and whether it can make equally accurate predictions in arbitrary classes of infinite partial wave amplitudes.
\end{itemize}
We will return to these issues in the next subsection.

\paragraph{Tests on higher-$L$ finite partial wave amplitudes.}

Another class of amplitudes outside the training-test dataset are the finite partial wave amplitudes with values of $L>3$. Exploring the quality of the NO predictions in this class shows that already at $L=4$ the NO fails to make any accurate predictions. This is, for example, apparent in the predictions presented at the bottom plot of Fig.\ \ref{fig:finite_L}, which depicts as a thick gray curve the exact result and as blue and orange dots the predictions of two separately trained NOs. The corresponding input modulus $B(z)$ appears on the top left plot of Fig.\ \ref{fig:finite_L}.

This case demonstrates that with the above-mentioned training on $L=1,2,3$ amplitudes the NO cannot fully reconstruct the unitarity equation, which would allow for valid predictions with arbitrary input modulus $B(z)$. It has been able to learn non-trivial elements of the unitarity constraints, but not all the information that these constraints entail.

\subsection{Neural Operator Setup II: Learning False Predictions}

The above observations raise the following related questions:
\begin{itemize}
    \item[$(a)$] Can NOs learn to rate the quality of their predictions producing reliable results without any reference to the unitarity equation \eqref{unitarity}?
    \item[$(b)$] Can NOs distinguish between  moduli $B(z)$ that are allowed by unitarity and moduli that are not?
        \item[$(c)$] Can NOs uncover quantifiable elements of the unitarity equation without having access to it?
\end{itemize}
In this subsection we want to focus exclusively on results that can be obtained without any use of the unitarity equation. This immediately removes PINNs as a viable methodology. In general, asking whether we can obtain any results without the underlying equation is interesting, because there is a plethora of problems in Physics and Mathematics where knowledge of the underlying structure is missing.

In the setup of Section~\ref{noset1}, the NOs are designed to make a prediction for arbitrary input modulus $B(z)$. Without prior knowledge of the unitarity equation \eqref{unitarity} it is impossible to deduce whether the solution exists, whether a prediction is valid, or to rate the quality of a prediction for a solution that exists. To address this difficulty, we propose setting up a slight variant of the NO of Section \ref{noset1}, where the output has two components: the predicted $\sin\phi(z)$ and a classifying label that we call {\it fidelity index} $\FF$, which contains information about the validity of the prediction. Accordingly, we now train the NO on two types of $(\vec B, \overrightarrow{\sin\phi})$ samples: the first contains the moduli and phases of valid finite partial wave amplitudes, and the second false moduli and phases that do not correspond to any amplitude. This setup should allow the NO to learn what it means to make a right prediction. 

\subsubsection{Hyperparameters and Training}

\paragraph{Hyperparameters.}
The results presented in this section  were obtained with a 1D TFNO that has the same Neural Network and optimization hyperparameters as the model in Section~\ref{hyperparams1}. However, in this case a different tensorization approach yields a larger model with a total number of 874,241 parameters (an order of magnitude larger than the one in the previous model of Section~\ref{hyperparams1}).

\paragraph{Training.}
We are using the same $z$-grid as in Section~\ref{hyperparams1} with 100 collocation points. The output vector of the NO is therefore 101-dimensional, including an extra element $v_{101}$ characterizing the validity of the output. In our runs we chose to train by assigning the value 10 to valid input-output pairs and -10 to false pairs. The fidelity index was defined as $\FF := \frac{10+v_{101}}{20}$, which assigns 1 to valid pairs and 0 to invalid ones.

We explored the results of training for a variety of datasets with varying fractions of true and false inputs-outputs. As one would expect, we observed that the quality of the classification output decreased when the fraction of false pairs was reduced. Here, we report results for a training-test dataset of 400K samples with the following composition: 75K true pairs for each of the $L=1,2,3$ amplitudes, and 175K false pairs. This yields a 43.75\% fraction of false samples. The inputs and outputs of the false samples were generated randomly from two different groups of $L=3$ amplitudes. We reserved 1.2K samples for testing and the samples were randomly mixed to put the true and false pairs in random order. With these specifications, we trained 56 independent NOs for 1.5K epochs.

\subsubsection{Tests and Observations}

\paragraph{Tests within the training-test dataset.}
The accuracy of the fidelity-index prediction can be probed by computing the difference 
\beq
\label{fidtest1aa}
\Delta \FF({\rm sample}) : = \big|\FF_{\rm pred}({\rm sample}) - \FF_{\rm ground-truth}({\rm sample})\big|
\eeq
between the predicted fidelity index and its ground truth, for each sample in the test dataset of 1.2K samples and separately for each of the above 56 trained NOs. Assuming that a prediction is considered correct when
\beq
\label{fidtest1ab}
\Delta \FF < C
\eeq
for some arbitrarily chosen $C<1$, we can go through the samples and register the number of times the inequality \eqref{fidtest1ab} is satisfied. This produces a success ratio $S_i(C)$ for the $i$-th NO. We can further average this success ratio over the NOs; we call the corresponding quantity $\bar S_\FF(C)$. For $C=0.2$ and 0.3 we find
\beq
\label{fidtest1ac}
\bar S_\FF(0.2) \simeq 73.94\% ~, ~~
\bar S_\FF(0.3) \simeq 75.34\%
~.
\eeq
The values $S_\FF^{(i)}(0.2)$ and $S_\FF^{(i)}(0.3)$ for the individual NOs are very close to the above averages. In other words, there is little variation between the different NOs in this datum. This suggests that (with the above cutoffs for $\Delta\FF$) the fidelity index makes the right classification roughly $75\%$ of the time, which is an encouraging sign of classification capacity but not an impressively high percentage.

We can also rephrase the above test in terms of a mean fidelity index $\bar \FF({\rm sample})$, which is defined for each sample as the average over the corresponding fidelity indices of all trained NOs. We can then define the difference $\Delta {\bar \FF}({\rm sample})$ as in \eqref{fidtest1aa} by replacing $\FF$ with $\bar \FF$, placing a cutoff as in \eqref{fidtest1ab} and computing the average $\bar S_{\bar \FF}(C)$ over the samples. For this quantity we find
\beq
\label{fidtest1ad}
\bar S_{\bar \FF}(0.2) \simeq 67.42\%~, ~~
\bar S_{\bar \FF}(0.3) \simeq 73\%
~,
\eeq
which is comparable to the previous result. We conclude that the average of the fidelity index over the NOs did not improve the classification capacity in this context.

These observations provide useful information about the performance of the NO as a classifier, but do not tell the whole story. In particular, we would now like to argue that the above tests do not really address some important aspects of the classification performance. Indeed, when one uses a NO to make a prediction for a never-before-seen modulus, it is very useful to know whether a predicted phase truly exists and can be considered correct with confidence, given an appropriately high fidelity index. Everything outside a small range of high fidelity values near 1 can be considered either as plausibly false for an existing phase, or false because a phase does not exist. This viewpoint rephrases the way we should measure the success ratio in the test dataset. 

Accordingly, we can now perform the following test. For each individually trained NO, we scan through the test dataset and count how many times the NO falsely affirms that the prediction is correct. The criterion for a prediction to be declared correct is
\beq
\label{fidtest1ae}
\big|\FF_{\rm pred}({\rm sample}) - 1 \big| < C
~.
\eeq
As we scan through the samples we count the cases where this inequality is satisfied and the ground-truth fidelity index vanishes (namely, the sample is false). That gives a percentage\footnote{We define the ratio that gives this percentage as the number of false predictions satisfying \eqref{fidtest1ae} divided by the total number of predictions satisfying \eqref{fidtest1ae}.} of failure $f^{(i)}(C)$ for the $i$-th NO. We want to examine if we can choose a small enough $C$ in \eqref{fidtest1ae} that yields high confidence in true predictions (that is, small $f^{(i)}(C)$), but we also want to check how many true cases we missed with this criterion. We can also compute the average over the NOs
\beq
\label{fidtest1af}
\bar f_{\FF}(C) = \frac{1}{N_{ops}}\sum_i f^{(i)}(C)
~,
\eeq
where $N_{ops}=56$ is the number of NOs. The label $\FF$ in $\bar f_{\FF}(C)$ is there to remind us that we are using the fidelity index of the individual NOs to evaluate the criterion \eqref{fidtest1ae} (this will change in a moment). As above, we did not observe significant variation in $f^{(i)}(C)$ among different NOs. Therefore, we quote here the values of $\bar f_{\FF}(C)$ for $C=0.01, 0.02, 0.05, 0.1, 0.2$ along with the fraction of correct predictions of true samples over the total number of true samples
\bea
\label{fidtest1ag}
\bar f_{\FF}(0.01) &=& 4.08 \%~, ~~~~
\frac{\rm correct~true~predictions}{\rm total~\#~of~true~samples} = 66.5\%
\nonumber\\
\bar f_{\FF}(0.02) &=& 5.37 \%~, ~~~~
\frac{\rm correct~true~predictions}{\rm total~\#~of~true~samples} = 67.3\%
\nonumber\\
\bar f_{\FF}(0.05) &=& 8.00 \%~, ~~~~
\frac{\rm correct~true~predictions}{\rm total~\#~of~true~samples} = 68.3\%
\\
\bar f_{\FF}(0.10) &=& 9.99 \%~, ~~~~
\frac{\rm correct~true~predictions}{\rm total~\#~of~true~samples} = 69.2\%
\nonumber\\
\bar f_{\FF}(0.20) &=& 11.71 \%~, ~~~
\frac{\rm correct~true~predictions}{\rm total~\#~of~true~samples} = 69.8\%
\nonumber
\eea

We notice that the predictions of a true solution are only 4.08\% of the times wrong when the fidelity index is inside the interval $[0.99, 1.01]$. This implies relatively high confidence in such predictions. We also notice that this criterion captures 66.5\% of the total number of true samples in the test dataset. As we increase $C$ (and with it the corresponding range of accepted fidelity indices) the fraction of wrong predictions increases and our confidence goes down, but the fraction of correct true predictions saturates. This implies that a small value of $C$ at the level of 0.01 is a preferable choice.

It is also interesting to re-evaluate these numbers using the mean fidelity index $\bar \FF$. In that case, we are first averaging the fidelity index over the trained NOs for a given sample to produce the corresponding mean fidelity index $\bar \FF_{\rm pred}({\rm sample})$, then we use it to impose a criterion like \eqref{fidtest1ae} and accordingly count which of the allowed samples are false predictions. This procedure yields a percentage of failure $f_{\bar \FF}(C)$ for the ``mean NO'' and the analog of \eqref{fidtest1ag} is 
\bea
\label{fidtest1aga}
f_{\bar \FF}(0.01) &=& 1.42 \%~, ~~~~
\frac{\rm correct~true~predictions}{\rm total~\#~of~true~samples} = 61.6\%
\nonumber\\
f_{\bar \FF}(0.02) &=& 1.56 \%~, ~~~~
\frac{\rm correct~true~predictions}{\rm total~\#~of~true~samples} = 65.2\%
\nonumber\\
f_{\bar \FF}(0.05) &=& 1.97 \%~, ~~~~
\frac{\rm correct~true~predictions}{\rm total~\#~of~true~samples} = 66.4\%
\\
f_{\bar \FF}(0.10) &=& 2.58 \%~, ~~~~
\frac{\rm correct~true~predictions}{\rm total~\#~of~true~samples} = 67.0\%
\nonumber\\
f_{\bar \FF}(0.20) &=& 4.63 \%~, ~~~~
\frac{\rm correct~true~predictions}{\rm total~\#~of~true~samples} = 67.1\%
\nonumber
\eea

We observe that the mean fidelity index produces a lower percentage of failure at the same value of $C$ (compared to the index of individual NOs) and, therefore, can be used to make predictions of correct phases with greater confidence. For example, the percentage of failed true predictions for the mean fidelity index at $C=0.01$ is only 1.42\%, compared to 4.08\% of the individual fidelity indices. The fraction of correct true predictions is comparable in both cases meaning that the mean fidelity index continues to detect essentially the same number of true samples with higher confidence.

\begin{figure}[t!]
    \begin{center}
        \includegraphics[width=0.495\textwidth]{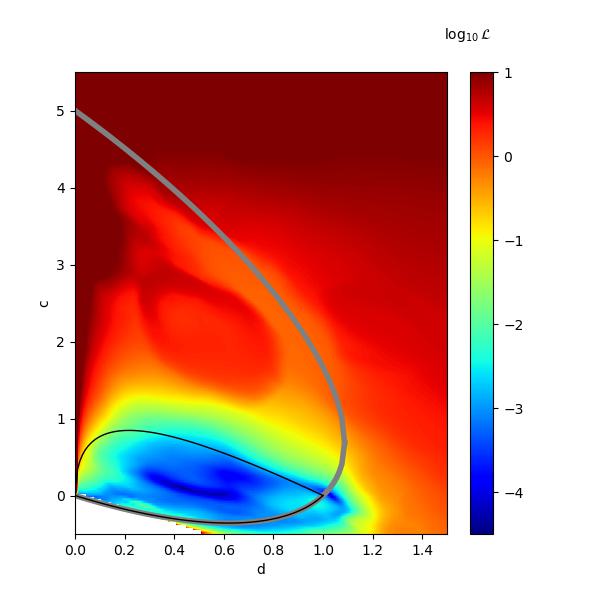}
        \includegraphics[width=0.495\textwidth]{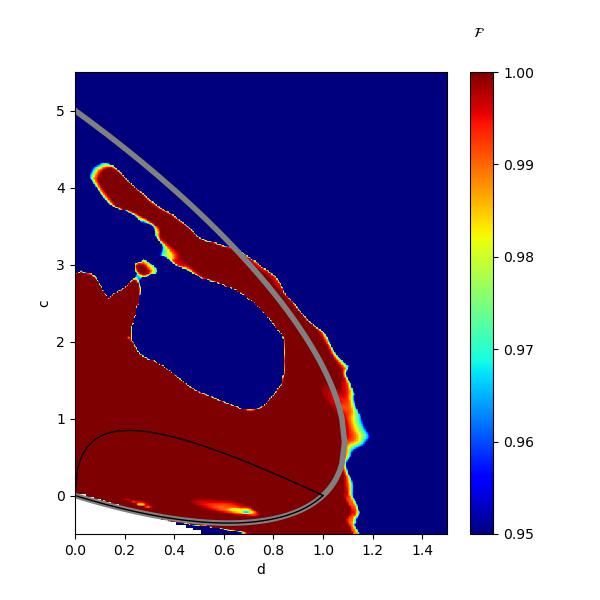}
    \end{center}
    \caption{\small Predictions for quadratic moduli $B(z)=cz^2+d$ by one of the 56 NOs trained on both true and false S-matrix phases. The heatmap on the left presents the $\log_{10}$ unitarity loss of the predictions. The heatmap on the right presents the value of the fidelity index $\FF$. The colorbar scale for the latter focuses on values between 0.95 and 1. Values below 0.95 are depicted in deep blue and values above 1 are depicted in deep red. As in previous plots, we have included the curve at $\sin\mu=1$ (light black) and the curve of the dual bound (thick gray).}
    \label{fig:heatmaps_class_example}
\end{figure}

\paragraph{Correlations between the unitarity loss and the fidelity index.}
As a further calibrating question we can ask whether the fidelity index correlates with the values of the unitarity loss. As an example, in Fig.\ \ref{fig:heatmaps_class_example} we contrast the heatmap of the $\log_{10}$ unitarity loss and the heatmap of the fidelity index for the predictions of one of the 56 trained NOs on the quadratic moduli $B(z)=cz^2+d$. There is visible correlation between the plots, and this is typical in many training runs both for linear and quadratic moduli. It is difficult, however, to make a precise quantitative statement about their relation. 

We also notice a clean separation between the predictions with high fidelity index (above 0.99) and predictions with low fidelity index (below 0.95). This is an interesting feature that correlates well with the above-mentioned observations about the fidelity index and its success rate. Unlike the unitarity loss, which varies smoothly between true and false predictions, the fidelity index appears to provide a more sharp acceptance/rejection criterion.

The results of the 56 trained NOs warrant some additional observations. First, we notice that the presence of the extra label that classifies the sample as true or false, has affected the nature of the predictions across the landscape of input moduli. This is visible in the comparison of the unitarity losses in the top right heatmap of Fig.\ \ref{fig:heatmaps} against the heatmap on the left of Fig.\ \ref{fig:heatmaps_class_example}. 

Second, the unitarity losses for the predictions of the 56 NOs that included the  fidelity label are typically slightly higher than those for the predictions of the NOs in Section~\ref{test1}, which did not involve any training on false samples. This is expected, since we were training with 300K true samples in Section~\ref{test1}, whereas the training here involves a smaller number of true samples, 225K.

Third, across the set of the 56 different NOs, we observed significant variation in the heatmaps of the predicted phases and their fidelity index. This observation hints at the complexities of the training process in this context and makes it harder to extract invariant information from individual training runs. It is therefore interesting to explore whether we can obtain information independent of the fluctuations of individual training iterations, reflecting real properties of the system, by collecting statistics from multiple runs.

\begin{figure}[t!]
    \begin{center}
        \includegraphics[width=0.495\textwidth]{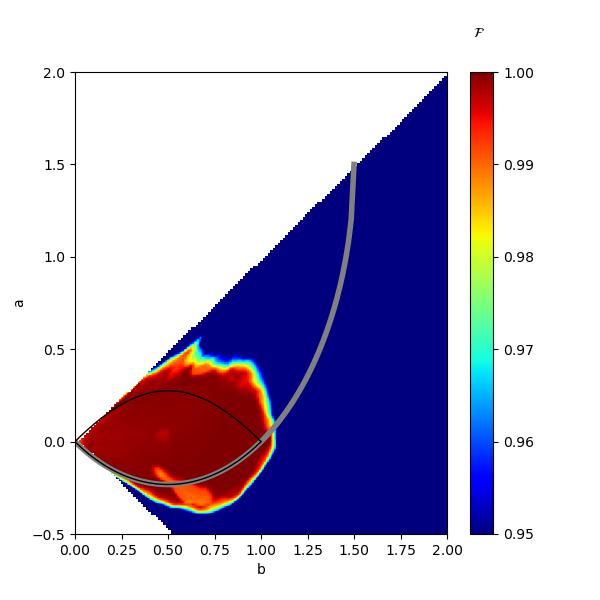}
        \includegraphics[width=0.495\textwidth]{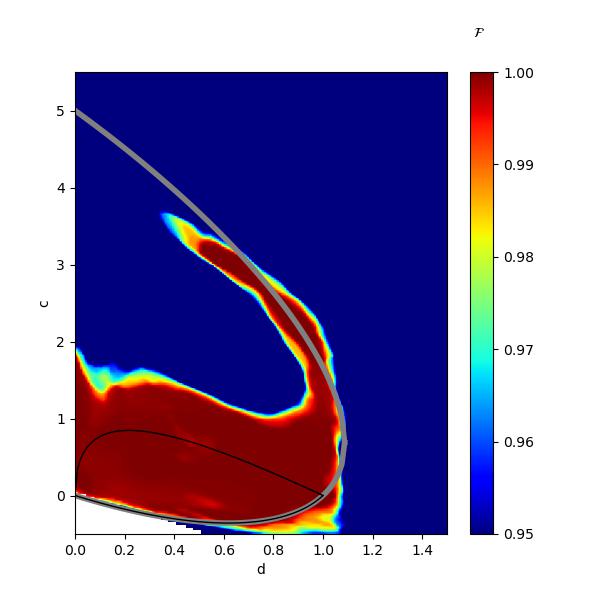}
    \end{center}
    \caption{\small The left heatmap depicts the value of the mean fidelity index $\bar \FF$ on the landscape of linear moduli $B(z)=az+b$. The right heatmap depicts the corresponding values of the mean fidelity index for quadratic moduli $B(z)=cz^2+d$. Notice that the colorbar scale focuses on values between 0.95 and 1. We have also included the curve at $\sin\mu=1$ (light black) and the curve of the dual bound (thick gray).}
    \label{fig:heatmaps_class}
\end{figure}

\paragraph{Performance of the mean fidelity index.}
The above discussion is further motivation in favor of the use of the mean fidelity index $\bar \FF$, which involves an average over independent NOs. In what follows, we evaluate $\bar \FF$ across the 56 previously trained NOs and plot the results in Fig.\ \ref{fig:heatmaps_class} across the landscape of linear and quadratic moduli. We observed that as we incorporated more and more NOs into the mean, there was an apparent convergence to the heatmaps of Fig.\ \ref{fig:heatmaps_class}. In the process, random fluctuating patterns from individual runs disappeared. 

For both linear and quadratic moduli we notice that the averaging over NOs preserves the sharp transition between high and low fidelity indices that was characteristic in individual runs. Additionally, we once again observe that the high fidelity index regions in red (with values above 0.99) match well with the expectations from the test-dataset for high confidence true predictions based on the mean fidelity index in this range.

For linear (quadratic) moduli, $\bar \FF$ is plotted on the left (right) heatmap of Fig.\ \ref{fig:heatmaps_class}, together with the $\sin\mu=1$ and dual bounds. We notice the characteristic concentration of high fidelity values around the $\sin\mu < 1$ region, which indicates that in the vicinity of this region the NOs correctly recognize predictions with the expected qualitative features of valid solutions. 

The heatmap of $\bar \FF$ in the quadratic scan exhibits some additional intriguing features that seem to fit well with features of the heatmap derived in Fig.\ 5 of Ref.\ \cite{Dersy:2023job} using PINNs. The high fidelity red region in our plot stretches above the upper $\sin\mu=1$ boundary in a manner that seems to correlate with a region of relatively low loss solutions (at the orders of $O(10^{-4.5})$ to $O(10^{-4})$) detected by PINNs.\footnote{In the heatmap of Fig.\ 5 in \cite{Dersy:2023job} the unitarity losses inside the $\sin\mu=1$ contour are of the order of $10^{-8}$. Here, we are referring to the losses of the PINN solutions right above that region and below the dual bound.} In addition, our heatmap has a characteristic upward tail that trails closely the dual bound in the vicinity of two points: one at $(c,d)\simeq (3,0.65)$ and another at $(c,d)\simeq (2.4, 0.9)$. The first point is tantalizingly close to the values $(c,d)=\sqrt{\frac{3}{8}} (5,1)\sim (3.06,0.61)$ of one of the finite-partial wave solutions with $L=2$ that the PINNs detect. The second point is similarly close to the values $(c,d)=\frac{5}{4}\sqrt{\frac{3}{7}} (3,1)\sim (2.45,0.82)$ of the second finite-partial wave solution with $L=2$ that the PINNs detect. Near the second point, our red region violates slightly the dual bound and so does a similar yellow blob in Fig.\ 5 from Ref.\ \cite{Dersy:2023job}. Interestingly, the NO of Section~\ref{test1}, without the mean fidelity index,  was unable to detect these features in the quadratic heatmap of Fig.\ \ref{fig:heatmaps}, but the NOs with the mean fidelity index seem to have picked them up. That is another indication that $\bar \FF$ is a promising measure for the detection of real features learned by the NOs.  

\begin{figure}[t!]
    \begin{center}
        \includegraphics[width=0.495\textwidth]{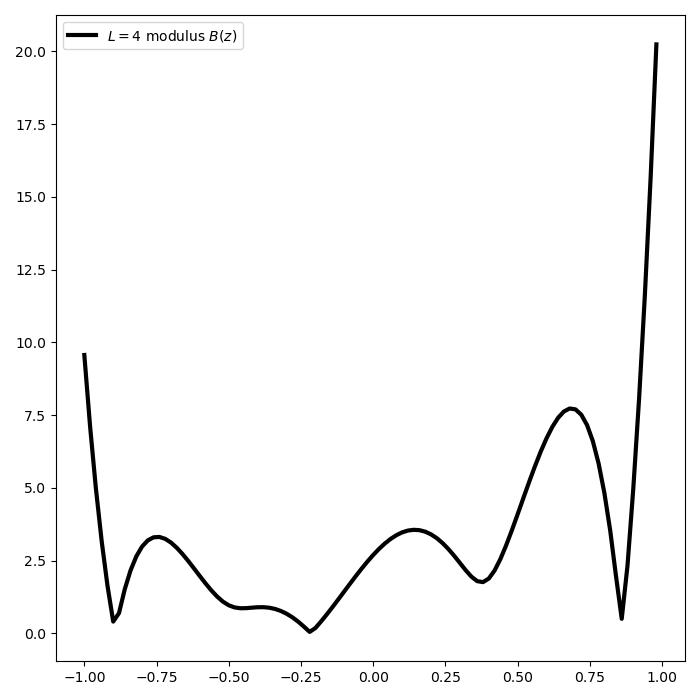}
        \includegraphics[width=0.495\textwidth]{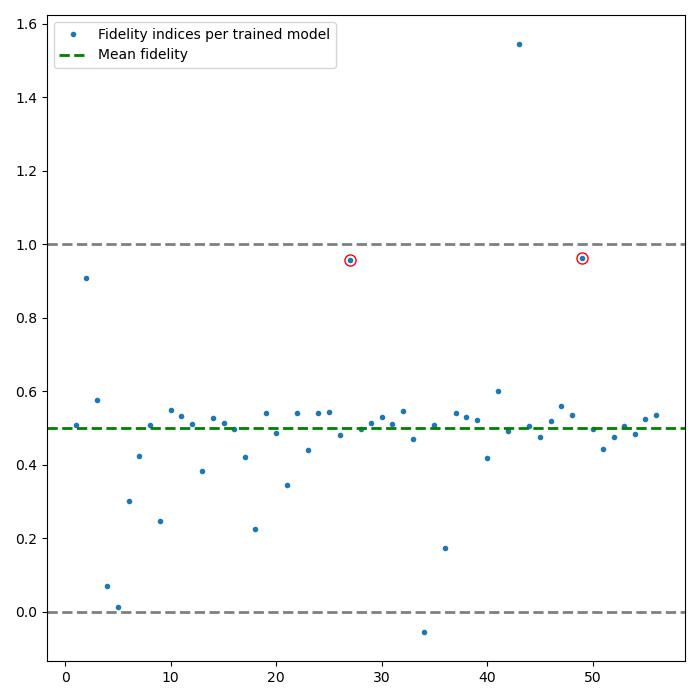}
        \includegraphics[width=0.495\textwidth]{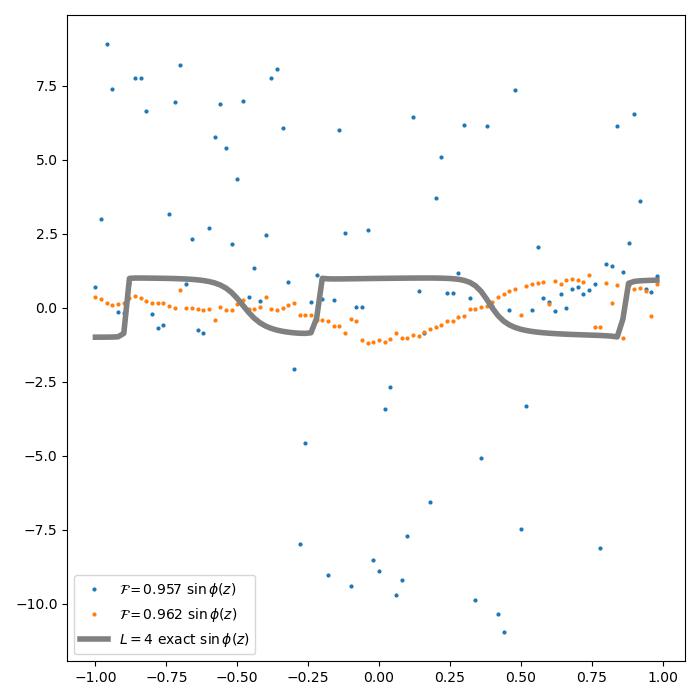} \vspace{0.3cm}
    \end{center}
    \caption{\small The top left plot depicts the modulus of a random $L=4$ amplitude. In the top right plot we present the fidelity indices for each of the 56 trained NOs evaluated on this specific modulus. The two points with a red circle around them represent the two predictions with fidelity indices closest to 1 (the actual values being 0.957 and 0.962, respectively). In the bottom plot we present the corresponding $\sin\phi(z)$ predictions of these two cases (in blue and orange) against the exact result represented by the thick gray curve.}
    \label{fig:finite_L}
\end{figure}

\paragraph{Detecting false predictions.}
We provided evidence that a high fidelity index (in the interval $[0.99,1.01]$) can confidently assess that the prediction is a valid solution. Outside this interval the validity of the prediction is less clear, but it is natural to expect that a low fidelity index will be associated more frequently with a false prediction. We would next like to examine more closely how the fidelity index behaves in situations where the NO fails. For that purpose, we return to the finite partial wave amplitudes with $L=4$, which proved to be a challenge for the NOs of Section~\ref{test1}. 

In Fig.\ \ref{fig:finite_L} we plot the fidelity index and some of the actual predictions from the 56 NOs trained on a combination of true and false samples. The NOs have been evaluated on the modulus of a randomly chosen $L=4$ amplitude, which is depicted on the top left plot of Fig.\ \ref{fig:finite_L}. On the top right plot, we present the values of the fidelity index for each of the trained NOs. The vast majority of the NOs exhibit a low index with mean value 0.499; this is consistent with the fact that the NOs fail to correctly reproduce the corresponding phase, as one can explicitly check by plotting the predicted output for $\sin\phi(z)$ against the exact result. This is useful: the NOs fail to generalize in this case, but they recognize correctly that this is the case and provide a clear indication of that information in the output.

Looking closer at the fidelity indices for each of the 56 NOs in the top right plot of Fig.\ \ref{fig:finite_L}, we also notice that out of the 56 NOs only two have a fidelity index within the interval $[0.95, 1]$. They are denoted with a red circle in the top right plot of Fig.\ \ref{fig:finite_L}. These two points correspond to fidelity indices 0.957 and 0.962. According to the previous discussion, they lie outside the region that captures a confident true prediction. The predicted $\sin\phi(z)$ for these NOs are presented at the bottom plot of Fig.\ \ref{fig:finite_L} against the exact result, denoted by the thick gray curve, in blue and orange respectively. The orange prediction, which has the higher fidelity index 0.962 is clearly better and qualitatively closer to an acceptable solution. It is a smoother function within the interval $[-1,1]$ (as one would expect from the sine of a real function), in contrast to the blue prediction that is chaotic and outside the interval $[-1,1]$. The NO has recognized the importance of these features and has assigned a higher fidelity index to the orange prediction. In principle, it is impossible to exclude the orange prediction as false, but the fact that it stands as a clear outlier in the statistics of Fig.\ \ref{fig:finite_L}, and that the mean fidelity index is very low with small deviation, suggest that the orange prediction is likely false.  

In conclusion, the above observations support using NOs within a statistical framework. In general situations, we propose the following approach: When the mean fidelity index suggests that a prediction should be rejected (as in Fig.\ \ref{fig:finite_L}), it should be discarded as potentially false. When the mean fidelity index is high (within the interval [0.99, 1.01]), one should accept the prediction as correct with high probability and extract predictions using the point-wise average of the predicted functions across the collection of NOs. Useful information can also be extracted by the point-wise standard deviation of the predicted functions.\footnote{A similar statistical approach was also advocated in the optimization schemes of Refs.\ \cite{Kantor:2022epi,Niarchos:2023lot}. In that context, the average of independent stochastic optimization runs (especially, those based on Reinforcement Learning) was always observed to provide better approximations.}

\section{Ambiguous Phases}
\label{double}

The problem of $S$-matrix phases is interesting for an additional reason. So far we have operated under the assumption that for a given modulus $B(z)$ there is a unique solution for the phase $\phi(z)$ (up to trivial ambiguities), or that there is no solution at all. As we briefly reviewed in Section \ref{background}, there are also cases of finely-tuned moduli that admit a doubly-ambiguous phase. Such cases were studied by several papers in the 1960s and 1970s, and still lack a general complete classification. More recently, Ref.\ \cite{Dersy:2023job} revisited the construction of such solutions using the PINN approach. In this section, we would like to explore if we can detect the ambiguous solutions of infinite partial wave amplitudes by training NOs on unique and ambiguous solutions of finite partial wave amplitudes. For the training we are going to use the fully classified ambiguous amplitudes with $L=2, 3$ \cite{Crichton:66,Atkinson:1973je,Berends:1973mg}. Clearly, this task will be much more subtle and demanding, compared to the generic configurations we have been discussing so far.

\subsection{Brief Note on $L=2,3$ Amplitudes with Phase Ambiguities}
\label{berends}

To generate training samples for $L=2,3$ amplitudes with ambiguous phases, we used the classification developed in \cite{Berends:1973mg}. Here, we briefly review the relevant construction and note some minor discrepancies in the original paper \cite{Berends:1973mg}. 

The approach of \cite{Berends:1973mg} involves an alternative decomposition of the partial wave amplitude in terms of the forward scattering amplitude (at $\theta = 0$) as
\begin{align}\label{altamp}
    F(z)=F(1) \prod_{l=1}^L \frac{z - W_l}{1- W_l}\;.
\end{align}
In this representation, all possible amplitudes with the same modulus at fixed $L$ can be obtained by acting on \eqref{altamp} with the transformations \cite{Gersten:1969ae}
\begin{align}\label{xfms}
 S: \operatorname{Re} F(1) &\rightarrow-\operatorname{Re} F(1)~,\cr
 T_l:  W_l &\rightarrow W_l^*\;.
\end{align}
Combinations of the above symmetries are also allowed as long as they do not lead to phases that are trivially related by sending $\delta_\ell \to - \delta_\ell$. Defining the variables $\zeta_\ell := e^{2 i \delta_\ell}$, it is straightforward to equate \eqref{noset1ab} with \eqref{altamp} and solve for $\zeta_\ell(W_\ell)$. One can then look for ambiguous solutions for the phase shifts $\delta_\ell$ by requiring that: $i)$ the $|\zeta_\ell|$ are left invariant by the transformations \eqref{xfms} and  $ii)$ $|\zeta_\ell|=1$, which is equivalent to imposing that the scattering is elastic. For $L = 2,3$ this procedure leads to a real one-parameter family of two-fold ambiguous phases (that are not trivially related) for specific intervals on the real line, as reported in Tables 1 and 2 of \cite{Berends:1973mg}. 

More specifically, for $L=2$ the only independent transformation that does not lead to trivially-related ambiguous phases is $ST_1$. Following the above steps this recovers the real one-parameter family of two-fold ambiguous solutions of \cite{Atkinson:1973je}, including the Crichton ambiguity \cite{Crichton:66}. 

For $L=3$ the only independent transformations that do not lead to trivially-related ambiguous phases are $T_1$ and $ST_1$. Analyzing the various possibilities leads to two classes of two-fold ambiguous families of solutions arising for each of the $T_1$ and $ST_1$ transformations. In this context, we report the following disagreement with two of the expressions in \cite{Berends:1973mg}. We find:
\begin{align}
\cos \eta &= \frac{\frac{\frac{15}{2 x}+\frac{3}{7}}{|W_1|^2}+\frac{135 |H'|^2 x}{2}+\frac{8 x}{15}+\frac{45}{2 x}+8}{3 |H'| (4 x+30)}\;,\\
|W_1|^2 &=\frac{252 c_5^2}{5 c_7 \left(7 c_5^2+9  c_1 |A|^2\right)}-\frac{2 c_1+1}{5 c_5}+\frac{(2 c_1+1) c_5}{ c_7|A|^2}
\end{align}
for (A.5) and (A.14) respectively.\footnote{Note that in \cite{Berends:1973mg} what we call $W_l$ is denoted as $F_l$.} We are in agreement, however, with all other formulas, as well as the conclusions of the analysis of \cite{Berends:1973mg} as presented in their Tables 1 and 2.

\subsection{Neural Operators on the Double Cover}
\label{noset2}

In order to incorporate the possibility of amplitudes that have the same modulus and two inequivalent phases, we set up a 1D TFNO that takes a single input $B(z)$, but outputs two $\sin\phi(z)$. On the $z$-grid with 100 collocation points this implies that the output is a 200-dimensional vector, which concatenates the 100-dimensional vectors of the two predictions. When the prediction is unique, the concatenated vectors are identical. We will report results without a fidelity index, but that is a feature that can be readily incorporated in this discussion.

\subsubsection{Hyperparameters and Training}

\paragraph{Hyperparameters.} 
Following a simple grid search, we observed a significantly larger dependence of the results on the NO hyperparameters for this problem. In what follows, we will report results based on NOs with essentially the same hyperparameters as in Section~\ref{hyperparams1}. The only hyperparameters that differ are the number of projection channels (we chose 256 instead of 512) and the number of layers (we chose 6 instead of 4). The resulting model has 72,745 parameters. 

\begin{figure}[t!]
    \begin{center}
        \includegraphics[width=0.495\textwidth]{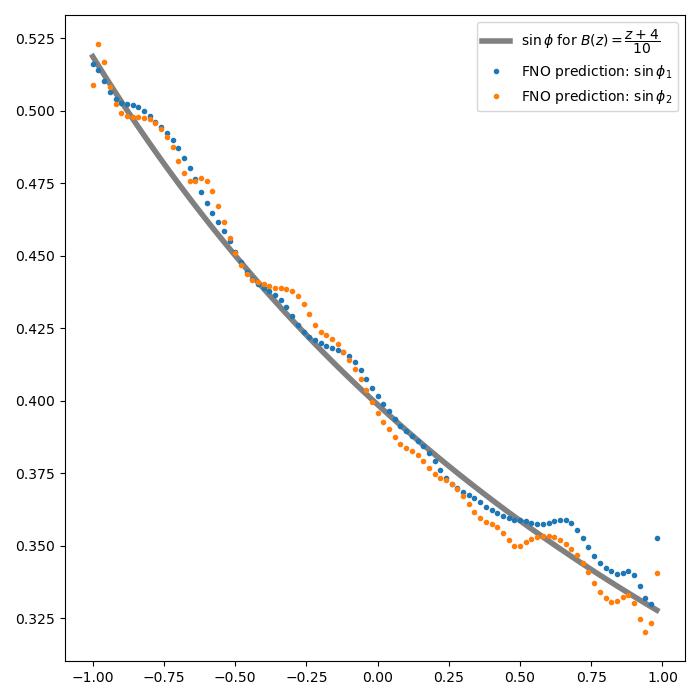}
        \includegraphics[width=0.495\textwidth]{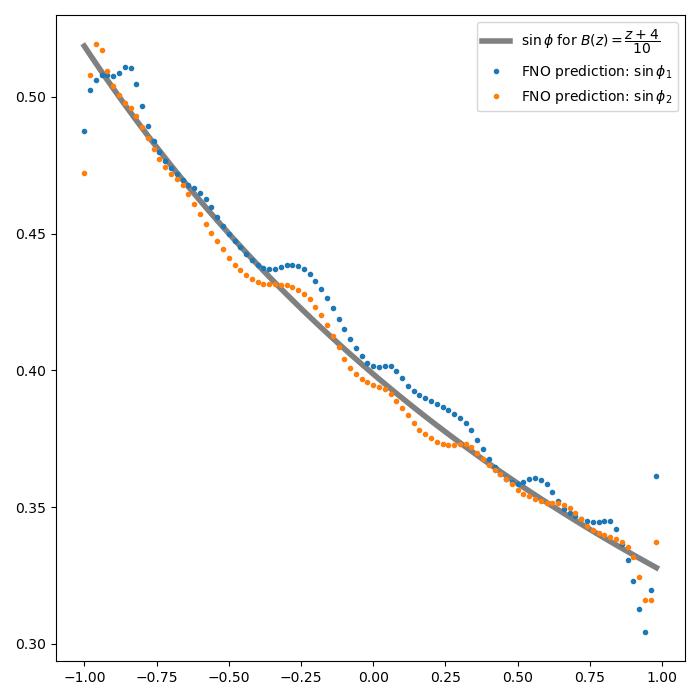} 
        \includegraphics[width=0.495\textwidth]{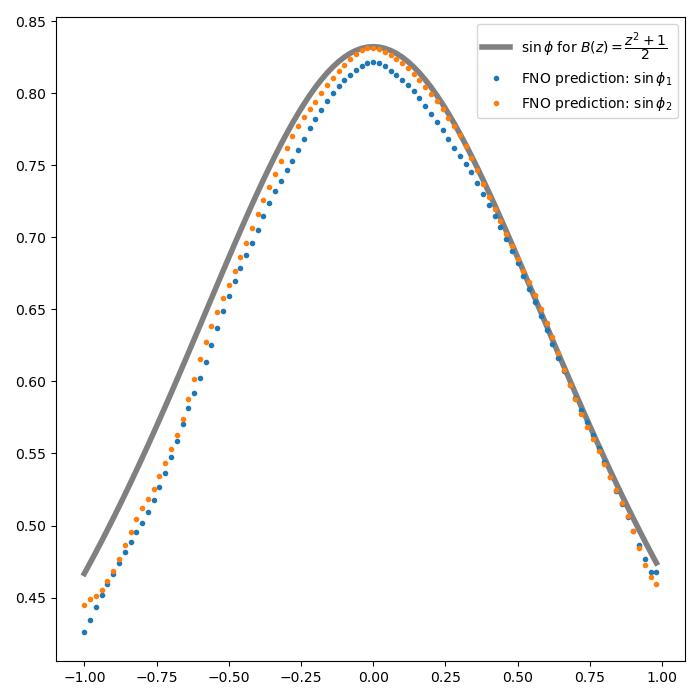}
        \includegraphics[width=0.495\textwidth]{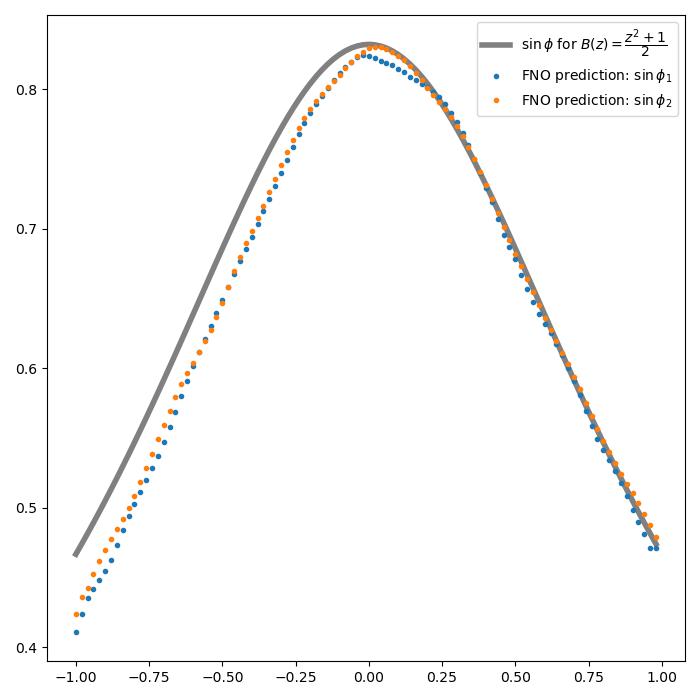}
        \hspace*{0.3cm}${\bf Neural~Operator~1}$ \hspace{4cm} ${\bf Neural~Operator~2}$
    \end{center}
    \vspace{0.3cm}
    \caption{\small The left column displays the exact $\sin\phi(z)$ (gray curve) and the two predictions of the first NO (blue and orange) for the linear and quadratic moduli $B(z)=\frac{1}{10}(z+4)$ and $B(z)=\frac{1}{2}(z^2+1)$. The right column displays the corresponding quantities for the second NO. Both NOs were trained on the same dataset and with the same hyperparameters.}
    \label{fig:linear_quad_double}
\end{figure}

\paragraph{Training.}
We attempted training with several types of datasets involving different ratios of unique and ambiguous solutions. Since we were limited to a relatively small range of amplitudes with ambiguous phases at $L=2,3$, we could not significantly increase the total number of samples, which in turn made the training less efficient. The results presented below are based on a dataset with a total number of 100K randomly chosen samples and the following split: 30K random $L=3$ amplitudes assumed to be unique, as well as 10K $L=2$ and 60K $L=3$ amplitudes with ambiguous phases sampled randomly across the different families of solutions summarized in the previous subsection. We trained on 99K of these samples and reserved 1K samples for testing. 

We present the results of two, independently-trained, NOs with the same hyperparameters, which were trained for 6.5K epochs.

\subsubsection{Tests and Observations}
\label{test2}

Once again, the NOs test well within the training-test dataset. Our purpose here is to explore whether they can achieve any sensible generalization outside their immediate training domain. We will not attempt an exhaustive analysis, opting instead for the study of a few examples for illustration purposes. Specifically, we will focus on the performance of: $a)$ the NOs on the linear and quadratic moduli of Fig.\ \ref{fig:single_predictions}---$B(z)=\frac{1}{10}(z+4)$ and $B(z)=\frac{1}{2}(z^2+1)$)---that have an infinite partial wave expansion and no phase ambiguities, and $b)$ one of the solutions with phase ambiguities in Ref.\ \cite{Atkinson:1977ia}---with parameter $z_1=\frac{6}{5}+\frac{3}{5}i$---that was also discussed in Ref.\ \cite{Dersy:2023job}; see e.g.\ Fig.\ 12 of that paper.

In Fig.\ \ref{fig:linear_quad_double} we present the predictions of the two NOs for the linear and quadratic moduli. In both cases, the two predicted phases are close to each other and close to the unique exact phase, but the accuracy of the results is obviously lower compared to the results of the previous sections. This is reasonable, since we only trained with 30K unique samples (compared to 300K unique samples in Section~\ref{test1}).

\begin{figure}[t!]
    \begin{center}
        \includegraphics[width=0.495\textwidth]{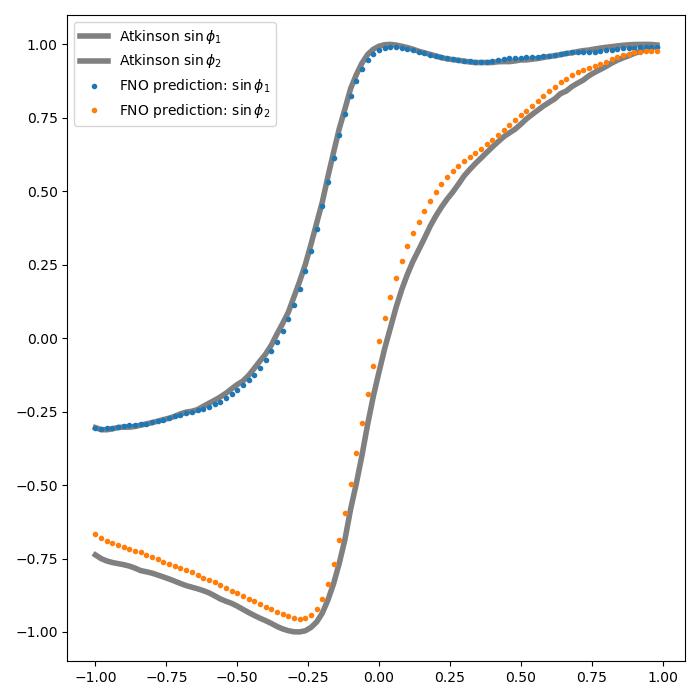}
        \includegraphics[width=0.495\textwidth]{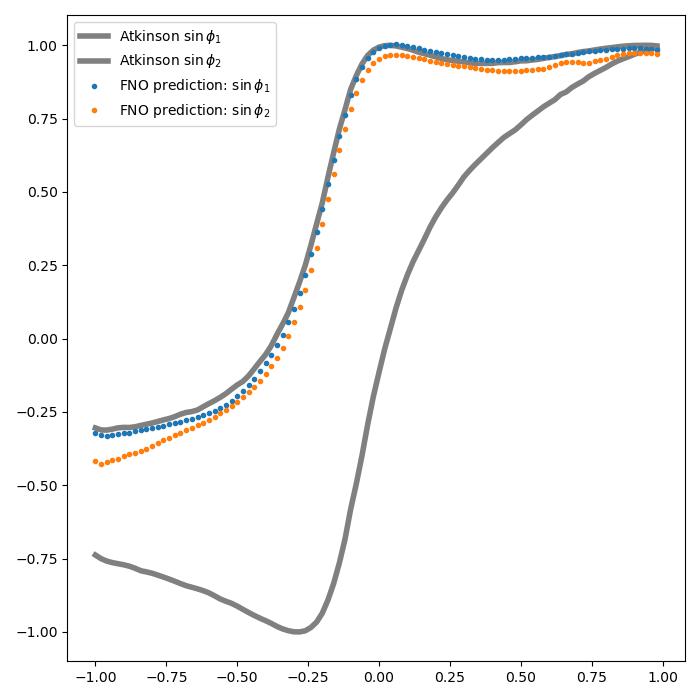}
        \hspace*{0.3cm}${\bf Neural~Operator~1}$ \hspace{4cm} ${\bf Neural~Operator~2}$
    \end{center}
    \caption{\small The left plot displays the two predictions of the first NO (blue and orange) against the solutions for $\sin{\phi(z)}$ for the Atkinson {\it et al.} modulus with $z_1=\frac{6}{5}+\frac{3}{5}i$. The right plot displays the predictions of the second NO. The first NO detected both solutions, while the second NO only one. Both NOs were trained on the same dataset, with the same hyperparameters and for the same number of epochs.}
    \label{fig:atkinson_double}
\end{figure}

In Fig.\ \ref{fig:atkinson_double} we display the corresponding predictions of the two NOs for the Atkinson {\it et al.} $z_1=\frac{6}{5}+\frac{3}{5}i$ modulus. The first NO detects both phases, but the second detects only one of them. More generally, over several runs we observed that properly trained NOs would see either one or both solutions. More frequently, they would detect only one solution (the same one that the second NO detects in Fig.\ \ref{fig:atkinson_double}).

We also tested the above NOs on the $z_1=0.31+0.95i$ ambiguous amplitude of Ref.\ \cite{Dersy:2023job} that has $\sin\mu\simeq 1.67$; see Fig.\ 15 in that paper. The NOs predicted a unique output partially approximating one of the solutions of Ref.\ \cite{Dersy:2023job} with low accuracy. We observed that the prediction was more sensitive (compared to other inputs) to the precise numerics of the input modulus. This is an expected difficulty in general, as it involves generalization to measure-zero configurations and we have no dynamical way to tune the input modulus. 

It would be interesting to explore if these problems can be addressed in the following manner: Still within the setting of not invoking the unitarity equation, one could first train a NO (or an ensemble of NOs) to produce a (mean) prediction of two $\sin\phi$ and a (mean) fidelity index. Then, in search of ambiguous solutions within the class where the NOs can generalize, one could run a PINN with a NN that models the modulus $B_\theta(z)$ (with $\theta$ the NN parameters) and a loss function that has two contributions: $i)$ a repulsive potential for the two $\sin\phi$ and $ii)$ a potential involving the (mean) fidelity index, e.g.\ of the form $(\FF-1)^2$. Both contributions are functionals of $B_\theta(z)$ and the idea would be to optimize the PINN parameters $\theta$ so that it produces moduli with two inequivalent phases and a high fidelity index close to 1.

\section{Conclusions and Outlook}\label{conclusions}

In this paper we used Fourier Neural Operators to study properties of amplitudes in elastic $2\to 2$ scattering processes. Unlike previous approaches, we did not invoke the unitarity equation \eqref{unitarity} to relate the modulus and phase, but tried to extract information about this relation from supervised training on random amplitudes with a finite partial wave expansion and $L=1,2,3$. We observed that NOs can generalize non-trivially outside this class, successfully recovering (after a single training process) the heatmaps of \cite{Dersy:2023job} for arbitrary linear and quadratic amplitude moduli. A similar approach was also applied to the two-fold ambiguous phase solutions. Even though this case is generically much harder, as it concerns subtle properties of finely-tuned configurations,
it was nevertheless possible to demonstrate in specific examples that the NOs can generalize to recover two inequivalent phases for amplitudes with infinite partial wave expansions. 

The question of how NOs generalize is not only central to this paper but also to the broader field of Artificial Intelligence. The answer can depend on many factors, which are usually hard to identify: the nature of the training dataset, the choice of hyperparameters and the details of the training, to name but a few. In the main text, we observed that within our specific setup the NO could learn several---but not all---non-trivial properties of the underlying general structure. For example, it could generalize to a class of amplitudes of infinite partial wave expansions, but failed on amplitudes with finite partial wave expansions for $L>3$. Moreover, by simply training on true modulus-phase pairs, the NO could not detect the cases where a modulus is inadmissible. For that reason, it was crucial to train on both true and false samples, which were distinguished by an extra classifying label that we called the fidelity index. It was clear from several examples that this index could extract useful information about properties of scattering amplitudes, hidden inside the (inaccessible) unitarity equation \eqref{unitarity}. We emphasized the importance of averaging over independent NOs and provided evidence that it can be used to increase the confidence of the predictions and reduce optimization noise during training, enabling us to isolate true system information. In particular, the mean fidelity index made the predictions more robust and allowed the NO to rate its own performance. 

We are excited by the potential use of similar approaches in other---possibly harder---problems, where the underlying structure is obscure, i.e. it is impossible to directly solve a system of equations or to directly compute relevant quantities. For instance, it would be interesting to explore whether objects similar to the fidelity index can be defined (using NOs or other Machine Learning algorithms, especially generative AI algorithms) for other systems. In addition, the examples presented in this paper seem to indicate that by studying the statistics of learners for the same training dataset and hyperparameters, one can distill information about what this particular class of algorithms can---and cannot---learn without recourse to the unknown microscopics, hence providing a new road towards structures we do not yet understand.

\section*{Acknowledgments}
The work of VN was partially supported by the H.F.R.I call “Basic research Financing (Horizontal support of all Sciences)” under the National Recovery and Resilience Plan “Greece 2.0” funded by the European Union – NextGenerationEU (H.F.R.I. Project Number: 15384). The work of CP was partially supported by the Science and Technology Facilities Council (STFC) Consolidated Grants ST/T000686/1 and ST/X00063X/1 “Amplitudes, Strings \& Duality”. Calculations were performed using the Sulis Tier 2 HPC platform hosted by the Scientific Computing Research Technology Platform at the University of Warwick. Sulis is funded by EPSRC Grant EP/T022108/1 and the HPC Midlands+ consortium.


\bibliography{NO}

\end{document}